\documentclass[longauth,traditabstract]{aa} 
\usepackage{graphicx}
\usepackage{txfonts}
\begin{document}
   \title{The CHESS chemical Herschel surveys of star forming regions: Peering into the protostellar shock
   L1157-B1}

   \subtitle{I. Shock chemical complexity\thanks{{\it Herschel} is an ESA space observatory
   with science instruments provided by European-led principal Investigator consortia and with 
   important partecipation from NASA}}

   \titlerunning{Peering into the protostellar shock
    L1157-B1: Shock Chemical Complexity}

   \author{
%
%
Codella C. \inst{1} \and 
Lefloch B. \inst{2} \and 
Ceccarelli C. \inst{2} \and 
Cernicharo J. \inst{3} \and 
Caux E. \inst{4} \and 
Lorenzani A. \inst{1} \and
Viti S. \inst{5,6} \and 
Hily-Blant P. \inst{2} \and 
Parise B. \inst{7} \and 
Maret S. \inst{2} \and
Nisini B. \inst{8} \and
Caselli P. \inst{9,1} \and
Cabrit S. \inst{10} \and
Pagani L. \inst{10} \and
Benedettini M. \inst{6} \and 
Boogert A. \inst{11} \and
Gueth F. \inst{12} \and
Melnick G. \inst{13} \and
Neufeld D. \inst{14} \and
Pacheco S. \inst{2} \and
Salez M. \inst{10} \and
Schuster K. \inst{12} \and
%
%
Bacmann A.\inst{2,15} \and
Baudry A. \inst{15} \and
Bell T. \inst{16} \and
Bergin E.A. \inst{17} \and
Blake G. \inst{16} \and
Bottinelli S. \inst{4} \and
Castets A. \inst{2} \and
Comito C. \inst{7} \and
Coutens A. \inst{4} \and
Crimier N. \inst{2,3} \and
Dominik C. \inst{18,19} \and
Demyk K. \inst{4} \and
Encrenaz P. \inst{10} \and
Falgarone E. \inst{10} \and
Fuente A. \inst{20} \and
Gerin M. \inst{10} \and
Goldsmith P. \inst{21} \and
Helmich F. \inst{22} \and
Hennebelle P. \inst{10} \and
Henning Th. \inst{23} \and
Herbst E. \inst{24} \and
Jacq T. \inst{15} \and
Kahane C. \inst{2} \and
Kama M. \inst{18} \and
Klotz A. \inst{2} \and
Langer W. \inst{21} \and
Lis D. \inst{16} \and
Lord S. \inst{16} \and
Pearson J. \inst{21} \and
Phillips T. \inst{16} \and
Saraceno P. \inst{6} \and
Schilke P. \inst{7,25} \and
Tielens X. \inst{26} \and
van der Tak F. \inst{22} \and
van der Wiel M. \inst{27,22} \and
Vastel C. \inst{4} \and
Wakelam V. \inst{15} \and
Walters A. \inst{4} \and
Wyrowski F. \inst{7} \and
Yorke H. \inst{21}  \and
%
%
Borys C. \inst{16} \and
Delorme Y. \inst{10} \and
Kramer C. \inst{28} \and
Larsson B. \inst{29} \and
Mehdi I. \inst{21} \and
Ossenkopf V. \inst{25} \and
Stutzki J. \inst{25} 
          }

   \institute{
INAF, Osservatorio Astrofisico di Arcetri, Firenze, Italy: \email{codella@arcetri.astro.it}
\and
Laboratoire d'Astrophysique de Grenoble, UMR 5571-CNRS, Universit\'e Joseph Fourier, Grenoble,
France
\and
Centro de Astrobiolog\`{\i}a, CSIC-INTA, Madrid, Spain
\and
CESR, Universit\'e Toulouse 3 and CNRS, Toulouse, France
\and
Department of Physics and Astronomy, University College London, London, UK
\and
INAF, Istituto di Fisica dello Spazio Interplanetario, Roma, Italy
\and
Max-Planck-Institut f\"{u}r Radioastronomie, Bonn, Germany
\and
INAF, Osservatorio Astronomico di Roma, Monte Porzio Catone, Italy
\and
School of Physics and Astronomy, University of Leeds, Leeds, UK
\and
Observatoire de Paris-Meudon, LERMA UMR CNRS 8112. Meudon, France
\and
Infared Processing and Analysis Center, Caltech, Pasadena, USA
\and
Institut de RadioAstronomie Millim\'etrique, Grenoble, France
\and
Center for Astrophysics, Cambridge MA, USA
\and
Johns Hopkins University, Baltimore MD, USA
\and
CNRS/INSU, Laboratoire d'Astrophysique de Bordeaux, Floirac, France
\and
California Institute of Technology, Pasadena CA, USA
\and
University of Michigan, Ann Arbor, MI 48109, USA
\and
Astronomical Institute 'Anton Pannekoek', University of Amsterdam, Amsterdam, The Netherlands
\and
Department of Astrophysics/IMAPP, Radboud University Nijmegen,  Nijmegen, The Netherlands
\and
IGN Observatorio Astron\'{o}mico Nacional, Alcal\'{a} de Henares, Spain
\and
Jet Propulsion Laboratory, Caltech, Pasadena, CA 91109, USA
\and
SRON, Institute for Space Research, Groningen, The Netherlands
\and
Max-Planck-Institut f\"{u}r Astronomie, Heidelberg, Germany
\and
Ohio State University, Columbus OH, USA
\and
Physikalisches Institut, Universit\"{a}t zu K\"{o}ln, K\"{o}ln, Germany
\and
Leiden Observatory, Leiden University, Leiden, The Netherlands
\and
Kapteyn Astronomical Institute, Groningen, The Netherlands
\and
Institut de RadioAstronomie Millim\'etrique, Granada, Spain 
\and
Department of Astronomy, Stockholm University, Stockholm, Sweden
          }

   \date{Received date; accepted date}

\abstract {We present the first results of the unbiased survey of the
L1157-B1 bow shock, obtained with HIFI in the framework of the key program
Chemical {\it Herschel} surveys of star forming regions (CHESS). The L1157
outflow is driven by a low-mass Class 0 protostar and is considered the prototype of the so-called
chemically active outflows. The bright blue-shifted bow shock B1
is the ideal laboratory for studying the link between the hot
($\sim$ 1000-2000 K) component traced by H$_2$ IR-emission and
the cold ($\sim$ 10--20 K) swept-up material. The
main aim is to trace the warm gas chemically enriched by the
passage of a shock and to infer the excitation conditions in
L1157-B1.  
A total of 27 lines are identified in the 555--636 GHz region, down to 
an average 3$\sigma$
level of 30 mK. The emission is dominated by CO(5--4) and
H$_2$O(1$_{\rm 10}$--1$_{\rm 01}$) transitions, as discussed by
Lefloch et al. in this volume.  Here we report on the
identification of lines from NH$_3$, H$_2$CO, CH$_3$OH, CS, HCN,
and HCO$^{+}$. The comparison between the profiles produced by molecules released from dust
mantles (NH$_3$, H$_2$CO, CH$_3$OH) and that of H$_2$O is
consistent with a scenario in which water is also formed in the
gas-phase in high-temperature regions where sputtering or
grain-grain collisions are not efficient. The high excitation
range of the observed tracers allows us to infer, for the first
time for these species, the existence of a warm ($\ge$ 200 K) gas
component coexisting in the B1 bow structure with the cold and
hot gas detected from ground.}
 
\keywords{ISM: individual objects: L1157 --- ISM: molecules --- stars:
formation}

   \maketitle
%

\section{Introduction}

A newborn protostar generates a fast and well collimated jet, possibly
surrounded by a wider angle wind. In turn, the ejected material drives 
(bow-)shocks travelling through the surrounding high-density medium
and traced by H$_2$ ro-vibrational lines at
excitation temperatures of around 2000 K.
As a consequence, slower and cold (10--20
K) molecular outflows are formed by swept-up material, usually traced by
CO.  Shocks heat the gas and trigger several processes such as
endothermic chemical reactions and ice grain mantle sublimation 
or sputtering. Several molecular species undergo significant  
enhancements in their abundances (see e.g., van Dishoeck \& Blake
\cite{vanblake}), as observed by observations at millimeter
wavelengths towards a number of outflows (Garay et al. \cite{garay};
Bachiller \& P\'erez Guti\'errez 1997, BP97 hereafter; J\o{}rgensen et
al. \cite{jorge}).  The link between the gas components at $\sim$ 10 K
and the hot 2000 K shocked component is crucial to understanding how the 
protostellar wind transfers momentum and energy back to
the ambient medium. In this context, the understanding of the chemical
composition of a typical molecular bow-shock is essential bevause it
represents a very powerful diagnostic tool for probing its physical
conditions.

\begin{figure}
\centering
\includegraphics[angle=-90,width=5cm]{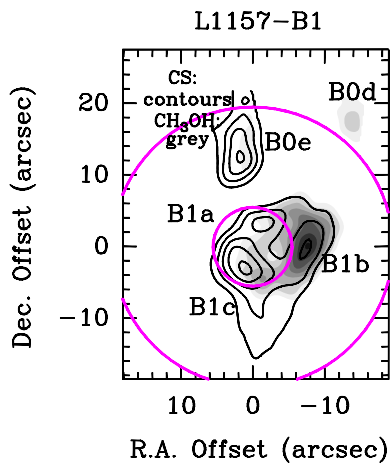}
\caption{The B1 clump. PdBI emission of
CH$_3$OH(2$_{\rm 1}$--1$_{\rm 1}$)A$^{-}$ (grey) on the CS(2--1)
one (contours), from BVC07. The maps are 
centred on the coordinates used for the present HIFI observations
$\alpha_{\rm J2000}$ = 20$^{\rm h}$ 39$^{\rm m}$ 10$\fs$2,
$\delta_{\rm J2000}$ = +68$\degr$ 01$\arcmin$ 10$\farcs$5, i.e. at
$\Delta\alpha$ = +25$\farcs$6 and $\Delta\delta$ = --63$\farcs$5
from the driving protostar. The labels
indicate the main B1 clumps detected in different tracers. 
Circles are for the HPBWs of the HIFI data presented here
(39$\arcsec$) and of Band 7 (11$\arcsec$), i.e., at the
highest frequencies of the CHESS surveys.}
\label{maps}
\end{figure}

The L1157 outflow, located at a distance estimated to be between 250 pc
(Looney et al. \cite{looney}) and 440 pc (Viotti \cite{viotti}) may be
regarded as the ideal laboratory for observing the effects of shocks on
the gas chemistry, being the archetype of the so-called chemically
rich outflows (Bachiller et al. \cite{bach01}).  The low-luminosity
(4--11 $L_{\rm \sun}$) Class 0 protostar IRAS20386+6751 drives a
precessing powerful molecular outflow associated with several bow
shocks seen in CO (Gueth et al. \cite{gueth96}) and in IR H$_2$
images (Davis \& Eisl\"offel \cite{davis}; Neufeld et
al. \cite{neufeld09}).  In particular, the brightest blue-shifted
bow-shock, called B1 (Fig. \ref{maps}), has been extensively mapped with
the PdB and VLA interferometers at mm- and cm-observations revealing a
rich and clumpy structure, the clumps being located at the wall of the
cavity with an arch-shape (Tafalla \& Bachiller
\cite{tafabachi}; Gueth et al. \cite{fred98}; Benedettini et
al. \cite{milena07}, hereafter BVC07; Codella et al. \cite{code09}).  L1157-B1 is well
traced by molecules thought to be released by dust mantles such as
H$_2$CO, CH$_3$OH, and NH$_3$ as well as typical tracers of
high-speed shocks such as SiO (e.g., Gusdorf et al. \cite{gus08b}). 
Temperatures $\simeq$ 60--200 K (from NH$_3$, CH$_3$CN, and SiO)
as well as around 1000 K (from H$_2$) have been derived (Tafalla \& Bachiller
\cite{tafabachi}; Codella et al. \cite{code09}; Nisini et al. \cite{nisini07}, in prep.).
However, a detailed study of the excitation conditions of the B1 structure 
has yet to be completed because of 
the limited range of excitation covered by the
observations performed so far at cm- and mm-wavelengths. Observations
of sub-mm lines with high excitation ($\ge$ 50--100 K above the ground
state) are thus required.

As part of the {\it Herschel} 
Key Program CHESS\footnote{http://www-laog.obs.ujf-grenoble.fr/heberges/chess/} 
(Chemical {\it Herschel} Surveys of Star forming regions),
L1157-B1 is currently being investigated with an unbiased spectral survey using
the HIFI instrument (de Graauw et al. \cite{hifi}). In this Letter, we
report the first results based on HIFI observations in the
555--636 GHz spectral window, confirming the chemical richness and
revealing different molecular components at different
excitation conditions coexisting in the B1 bow structure.
 
\section{Observations}

The observations were performed on 2009, August 1, during the
Performance Verification phase of the HIFI heterodyne instrument (de
Graauw et al. \cite{hifi}) on board of the {\it Herschel} Space Observatory
(Pilbratt et al. \cite{herschel}). The band called 1b (555.4--636.2 GHz)
was covered in double-sideband (DSB) with a total
integration time of 140 minutes. 
The Wide Band Spectrometer was used with a  
frequency resolution of 1 MHz. The typical HPBW is 
39$\arcsec$. The data were processed with the
ESA-supported package HIPE\footnote{HIPE is a joint development by the {\it Herschel} Science
Ground Segment Consortium, consisting of ESA, the NASA {\it Herschel} Science Center,
and the HIFI, PACS and
SPIRE consortia.} ({\it Herschel} Interactive Processing
Environment) for baseline subtraction and sideband
deconvolution and then analysed with
the GILDAS\footnote{http://www.iram.fr/IRAMFR/GILDAS} software. All the
spectra (here in units of antenna $T_{\rm a}$) were smoothed
to a velocity resolution of 1 km s$^{-1}$, except those showing the
weakest emission, which were smoothed to lower spectral resolutions
(up to 4 km s$^{-1}$). At a velocity resolution of 1 km s$^{-1}$, the rms noise is 6--13 mK 
($T_{\rm a}$ scale), depending on the line frequency. The main-beam
efficiency ($\eta_{mb}$) has not yet been reliably determined. 
When needed, we adopted an average $\eta_{mb}$ of 0.72.

\begin{table*}
\caption{List of molecular species and transitions observed with HIFI (Band 1b): 
CO and H$_2$O emission is discussed in Lefloch et al. (\cite{letter2}). Peak velocity and intensity
(in $T_{\rm a}$ scale), 
integrated intensity ($F_{\rm int}$), as well as the terminal velocities of the line emission 
($V_{\rm min}$ and $V_{\rm max}$) are reported.}
\label{lines}
\centering
\begin{tabular}{lrrrrcccc}
\hline
\multicolumn{1}{c}{Transition} &
\multicolumn{1}{c}{$\nu_{\rm 0}$$^a$} &
\multicolumn{1}{c}{$E_{\rm u}$$^a$} &
\multicolumn{1}{c}{$T_{\rm peak}$} &
\multicolumn{1}{c}{rms} &
\multicolumn{1}{c}{$V_{\rm peak}$} &
\multicolumn{1}{c}{$V_{\rm min}$} &
\multicolumn{1}{c}{$V_{\rm max}$} &
\multicolumn{1}{c}{$F_{\rm int}$} \\
\multicolumn{1}{c}{} &
\multicolumn{1}{c}{(MHz)} &
\multicolumn{1}{c}{(K)} &
\multicolumn{1}{c}{(mK)} &
\multicolumn{1}{c}{(mK)} &
\multicolumn{1}{c}{(km s$^{-1}$)} &
\multicolumn{1}{c}{(km s$^{-1}$)} &
\multicolumn{1}{c}{(km s$^{-1}$)} &
\multicolumn{1}{c}{(K km s$^{-1}$)} \\
\hline
o-H$_2$O(1$_{\rm 10}$--1$_{\rm 01}$) & 556936.002 & 27 & 910(17) & 17 & --0.37(1.00) & --25.4 & +7.6 &
11.68(0.10) \\
CH$_3$OH E ($11_{2,9}-10_{1,9}$) & 558344.500 & 168 & 47(10) & 10 & +0.60(1.00) & --2.9 & +2.7 &
0.16(0.02) \\
o-H$_2$CO(8$_{\rm 18}$--7$_{\rm 17}$) & 561899.318 & 118 & 92(4)$^b$ & 8 & +0.38(0.14)$^b$ & --4.5 & +3.6 &
0.49(0.03)$^b$ \\
CH$_3$OH E ($3_{2,2}-2_{1,2}$) & 568566.054 & 32 & 42(8) & 8 &  +0.60(1.00) & --2.6 & +2.7 &
0.31(0.02) \\
o-NH$_3$(1$_0$--0$_0$)     & 572498.068 &  28 & 122(7) & 7  & +1.03(0.80) & --6.9 & +5.0 & 0.89(0.03) \\
CO(5-4)                  & 576267.931 & 83 & 883(10) & 10 & +1.60(1.00) & --36.5 & +6.0 & 49.30(0.07) \\
p-H$_2$CO(8$_{\rm 08}$--7$_{\rm 07}$) & 576708.315 & 125 & 41(7) & 7 & +1.80(1.00)  & --5.4 & +2.7 &
0.22(0.02) \\
CH$_3$OH A$^{-}$ ($2_{2,1}-1_{1,0}$) & 579084.700 & 45 & 53(8) & 8 & +0.60(1.00) & --6.0 & +3.7 &
0.29(0.03) \\
CH$_3$OH E  ($12_{1,12}-11_{1,11}$) & 579151.003 & 178 & 44(8) & 8 & --0.30(1.00) & --4.0 & +4.0 &
0.25(0.02) \\
CH$_3$OH A$^{+}$ ($12_{0,12}-11_{0,11}$) & 579459.639 & 181 & 48(7) & 7 & +0.90(1.00) & --3.9 & +4.9 &
0.27(0.02) \\
CH$_3$OH A$^{+}$ ($2_{2,0}-1_{1,1}$) & 579921.342 & 45 & 42(7) & 7 & --0.60(1.00) &  --3.3 & +2.6 &
0.21(0.02) \\
CH$_3$OH E  ($12_{2,10}-11_{2,9}$) & 580902.721 & 195 & 11(4) & 4 & --3.00(3.00) &  --3.9 & +2.5 &
0.09(0.02) \\
CH$_3$OH A$^{+}$ ($6_{1,6}-5_{0,5}$) & 584449.896 & 63 & 88(9) & 9 & --0.30(1.00) &  --6.0 & +2.9 &
0.58(0.03) \\
CS(12--11)               & 587616.240 & 183 & 23(2)$^b$ & 5 & --0.61(0.57)$^b$  & --7.3 & +6.5 &
0.19(0.03)$^b$ \\
CH$_3$OH A$^{+}$ ($7_{3,5}-6_{2,4}$) & 590277.688 & 115 &  42(9) & 9 & +0.60(1.00)  &  --1.0 & +3.0 &
0.16(0.02) \\
CH$_3$OH A$^{-}$ ($7_{3,4}-6_{2,5}$) & 590440.291 & 115 &  40(9) & 9 & --0.60(1.00) &  --2.6 & +0.8 &
0.10(0.02) \\
CH$_3$OH E ($9_{0,9}-8_{1,8}$) & 590790.957 & 110 &  40(10) & 10 & +0.60(1.00) & --4.7  & +4.3 &
0.28(0.03) \\
o-H$_2$CO(8$_{\rm 17}$--7$_{\rm 16}$) & 600374.604 & 126 &  29(7)$^b$ & 9 & --0.14(0.57)$^b$  & --3.0 & +1.8
& 0.19(0.04)$^b$ \\
CH$_3$OH E ($4_{2,3}-3_{1,3}$) & 616979.984 & 41 &  47(6) & 6 & +0.90(1.00) & --4.5  & +2.6 &
0.26(0.02) \\
HCN(7--6)                & 620304.095 & 119 & 94(7) & 7  & --0.60(1.00) & --7.6  & +3.2 & 0.68(0.03) \\
HCO$^+$(7--6)            & 624208.180 & 119 & 30(3)$^b$ & 8 & +0.53(0.47)$^b$  & --3.6 & +4.5 &
0.11(0.03)$^b$ \\
CH$_3$OH A$^{-}$ ($3_{2,2}-2_{1,1}$) & 626626.302 & 52 &  18(5) & 5 & --1.20(4.00) & --2.5 & +6.3 &
0.18(0.07) \\
CH$_3$OH E ($13_{1,13}-12_{1,12}$) & 627170.503 & 209 & 33(7) & 7 & --0.30(1.00) & --3.2 & +2.5 &
0.15(0.02) \\
CH$_3$OH A$^{+}$  ($13_{0,13}-12_{0,12}$) & 627558.440 & 211 & 41(9) & 9 & +0.60(1.00)   & --3.6 & +2.5
& 0.19(0.02) \\
CH$_3$OH A$^{+}$  ($3_{1,2}-2_{1,2}$) & 629140.493 & 52 &   52(9) & 9 & +0.60(1.00)   & --3.5 & +3.7 &
0.24(0.03) \\
CH$_3$OH A$^{+}$  ($7_{1,7}-6_{0,6}$) & 629921.337 & 79 &   70(13) & 13 & +1.50(1.00)   & --3.9 & +3.7 &
0.37(0.04) \\
o-H$_2$CO(9$_{\rm 19}$--8$_{\rm 18}$) & 631702.813 & 149  &  66(4)$^b$ & 9 & +0.45(0.17)$^b$ & --1.5 & +2.6
& 0.22(0.02)$^b$ \\
\hline
\end{tabular}
\begin{center}
$^a$ Frequencies and spectroscopic parameters have been extracted from the Jet
Propulsion Laboratory molecular database (Pickett et al. \cite{pickett}) for all
the transition except those of CH$_3$OH, which have been extracted from the  
Cologne Database for Molecular Spectroscopy (M\"uller et al. \cite{muller}). Upper
level energies refer to the ground state of each symmetry. $^b$ Gaussian fit. \\
\end{center}
\end{table*}

\section{Different tracers at different velocities}

\begin{figure*}
\centering
\includegraphics[angle=-90,width=11cm]{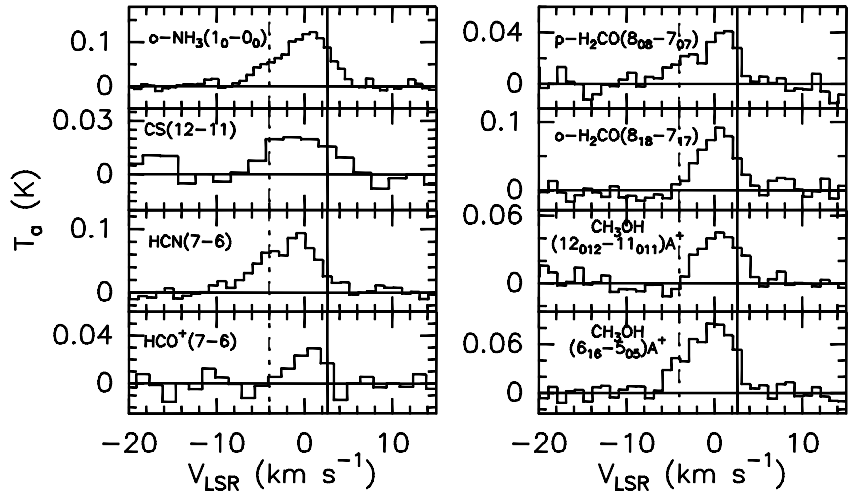}
\caption{Molecular line profiles observed towards L1157-B1: species
and transitions are reported in the panels. The vertical solid line
indicates the ambient LSR velocity (+2.6 km s$^{-1}$ from C$^{18}$O emission; BP97), while
the dashed one is for the secondary peak at --4.0 km s$^{-1}$.}
\label{spectra}
\end{figure*}

\begin{figure}
\centering
\includegraphics[angle=-90,width=7cm]{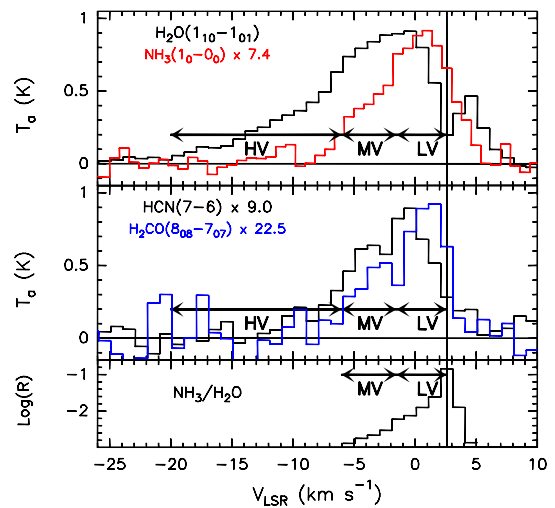}
\caption{{\it Top and Middle panel}: Comparison between the profiles of
NH$_3$(1$_0$--0$_0$), multiplied by a factor 7.4, H$_2$CO(8$_{\rm
17}$--7$_{\rm 16}$), multipled by a factor 22.5, HCN(7--6),  multipled
by a factor 9.0, and
H$_2$O(1$_{\rm 10}$--1$_{\rm 01}$), the latter from Lefloch et
al. (\cite{letter2}).  The vertical solid line indicates the
ambient LSR velocity (+2.6 km s$^{-1}$). The velocity ranges
arbitrarily defined as HV (--20,--6 km s$^{-1}$; traced by H$_2$O), MV (--6,
--1.5 km s$^{-1}$; outlined by the HCN and H$_2$CO secondary peak), and LV
(--1.5,+2.6 km s$^{-1}$; the rest of the blue wing) are drawn (see
text).
{\it Bottom panel}: Intensity NH$_3$/H$_2$O line ratio as a
function of velocity.}
\label{profiles}
\end{figure}

A total of 27 emission lines were detected, with
a wide range of
upper level energies, from a few tens to a few hundreds of Kelvin. Table 1 lists
the spectroscopic and observational parameters of all the transitions. 
For the first time, high excitation (up to
$\simeq$ 200 K) emission lines related to species whose abundance is
largely enhanced in shocked regions were detected. 
The CO(5--4) and H$_2$O(1$_{\rm
10}$--1$_{\rm 01}$) lines are analysed in Lefloch et
al. (\cite{letter2}).
Figure \ref{spectra} presents representative examples of line profiles
observed towards L1157-B1. All the spectra contain lines with blue-shifted wings
peaking near 0 km s$^{-1}$, which have a terminal velocity 
equal to $\sim$ --8,--6 km s$^{-1}$.  Previous PdBI observations
showed that L1157-B1 is associated with very high velocities
(HVs) of as low as $\simeq$ --20 km s$^{-1}$ ($v_{\rm LSR}$ = +2.6 km
s$^{-1}$, BP97). We cannot exclude the lack of detected emission in the HV
regime in the present HIFI spectra being caused by their 
relatively low signal-to-noise (S/N) ratio. The PdBI images indicate
that the brightness of the emission lines in the HV regime 
is indeed weaker than the emission at
low velocities by a factor of 5--10. The spectra in
Fig. \ref{spectra} clearly show that this weak emission would lie
below the noise. On the other hand, the HV gas is detected 
in the very bright lines of CO
and H$_2$O (Lefloch et al. \cite{letter2}).
We note that the HV emission is mostly confined to within the eastern B1a
clump (Fig. \ref{maps}), within an emitting region of size
$\le$ 10$\arcsec$ (Gueth et al. \cite{fred98}; BVC07), 
whereas low velocity lines originate in both 
the bow-structure and the walls of the outflow 
cavity (e.g., the B0e and B0d in Fig. \ref{maps}), 
of typical size 15$\arcsec$--18$\arcsec$. Therefore,
the forthcoming HIFI-CHESS observations at higher frequencies and 
higher spatial resolution (see the dashed circle in Fig. \ref{maps})
should allow us to study the HV wings in species other than CO and
H$_2$O.

The uniqueness of HIFI lies in its high spectral profile resolution 
for many high excitation transitions of a large number of molecular species. 
The analysis of the present HIFI spectra reveals a secondary 
peak occuring between --3.0 and --4.0 km s$^{-1}$ (here defined medium velocity, MV) 
and well outlined by e.g.,  
HCN(7--6). The MV peak is also visible in NH$_3$(1$_0$--0$_0$) and in
some lines of CH$_3$OH and H$_2$CO (see Fig. \ref{profiles}), but its occurrence does not show
any clear trend with the choice of tracer of line excitation.   
No single-dish spectra had previously detected this spectral feature
(BP97; Bachiller et al. \cite{bach01}). 
An inspection of the spectra
observed at PdBI shows that the MV secondary peak is observed in a
couple of lines of the CH$_3$OH(2$_{\rm K}$--1$_{\rm K}$) series (see
Fig. 3 of BVC07) and only towards the
western B1b clump (size $\sim$ 5$\arcsec$). This finding implies that 
there is a velocity component originating mainly in the western side of
B1, while the HV gas is emitted from the eastern one (see above).

Figure \ref{profiles} compares the profiles of
the NH$_3$(1$_0$--0$_0$) and H$_2$CO(8$_{\rm 17}$--7$_{\rm 16}$) lines
with the H$_2$O(1$_{\rm 10}$--1$_{\rm 01}$) profile, where the S/N
allows such an analysis (MV and LV ranges). 
By assuming that the emission in the MV range is 
optically thin (including the H$_2$O line) and originates in
the same region, we obtained from the comparison of their profiles a
straightforward estimate of the relative abundance ratios of the gas
at different velocities.
As a notable example, the NH$_3$/H$_2$O intensity ratio decreases
by a factor $\sim$ 5 moving towards higher velocities
(Fig. \ref{profiles}), implying that a similar decrease in the
abundance ratios occurs.  
This may reflect different pre-shock ice compositions in the
gas emitting the MV emission.
Alternatively, this behavior is consistent with 
NH$_3$ being released by grain mantles, but water both being released by
grain mantles and, in addition, copiously forming in the warm shocked gas from 
endothermic reactions, which convert all gaseous atomic oxygen into
water (Kaufman \& Neufeld 1997; Jim\'enez-Serra et
al. \cite{jimenez}, and references therein). The water
abundance may be enhanced with respect to ammonia in the fast and warm
($\geq 220$ K) gas, which might explain why the H$_2$O wings are larger
than those of NH$_3$, CH$_3$OH, and H$_2$CO, all species being directly
evaporated from dust grain mantles.

\section{Physical properties along the B1 bow shock}

\begin{figure}
\centering
\includegraphics[angle=-90,width=8cm]{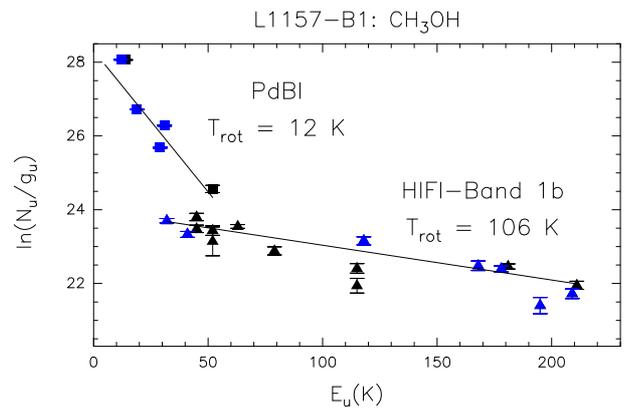}
\caption{Rotation diagrams for the CH$_3$OH transitions measured with
HIFI (triangles) and from ground (PdBI; squares). 
Black and blue points are for A- and E-form, respectively. 
The parameters $N_{\rm u}$, $g_{\rm u}$, and $E_{\rm u}$ are,
respectively, the column density, the degeneracy and the energy 
(with respect to the ground state of each symmetry) of the upper level.
The derived values of
the rotational temperature are reported: (i) 106 K, for the HIFI
lines covering the $E_{\rm u}$ = 32-211 K excitation range and (ii)
12 K (as Bachiller et al. \cite{bach95}), for the
PdBI lines, at lower excitation.}
\label{rotameta}
\end{figure}

We detected several lines from
CH$_3$OH (17 lines with upper level energies up to 211 K).
We can derive a first estimate of the emitting gas
temperature by means of the standard
analysis of the rotational diagram. 
We show the case of methanol (A- and E-forms) in
Fig. \ref{rotameta}. 
The derived rotational temperature ($T_{\rm rot}$) is 106 K (with
an uncertainty of $\sim$ 20 K), which represents a
lower limit to the kinetic temperature ($T_{\rm kin}$). In the same figure, we 
report the methanol lines (2$_{\rm K}$--1$_{\rm K}$) observed with
PdBI and whose intensity is integrated in the HIFI 39$\arcsec$ beam.
The $T_{\rm rot}$ derived from the ground-based data (based
only on lines with $E_{\rm u}$ $\le$ 50 K; BVC07) is definitely lower,
$\sim$12 K, in perfect agreement with that found with the 30-m
spectra in the same excitation range by Bachiller et
al. (\cite{bach95}). As discussed by Goldsmith \& Langer
(\cite{gold}), this behavior may be caused by either two components
at different temperatures or both non-LTE effects and line opacity.
These two possibilities cannot be distinguished based only on the
rotational diagram. 
However, given that a range of $T_{\rm kin}$ and $n_{\rm H_2}$ is naturally expected in a shock,
if we were to assume that two gas components provide an  
explanation, they would not only have different temperatures but
also a different column densities. Taking the filling factor ff =
0.13, derived by the CH$_3$OH maps obtained at the PdBI, the low
temperature component column density is 8 $\times$ 10$^{14}$
cm$^{-2}$ (in agreement with Bachiller et
al. \cite{bach95}), whereas the high temperature component has 
a column density of around 10$^{14}$ cm$^{-2}$. 
We note that the rotation diagrams obtained
for the MV and LV CH$_3$OH emission separately do not allow us
to infer any clear difference.

It is possible to more tightly constrain the emitting gas temperature and
$n_{\rm H_2}$ density for
the species where the collisional rate coefficients are known, 
by performing of a non-LTE
analysis. To this end, we used the non-LTE excitation code RADEX
with an escape probability formalism for the radiative transfer 
(Van der Tak et al. \cite{radex}) coupled with the LAMDA database
(Sch\"oier et al. \cite{lamda}). 
Methanol is the species detected in the largest number of lines.
The full non-LTE study will be reported in a forthcoming paper.
Here we analysed only the E-form, for which the 
collisional rate coefficients are available (Pottage et al. \cite{pottage}).
The major result of
this analysis is that for a range of densities of 10$^{3}$--10$^{7}$ cm$^{-3}$, 
the gas temperature exceeds 200 K.
A similar result is obtained by considering H$_2$CO emission.

Finally, by combining the HIFI CS(12--11)
line with CS(2--1) and (3--2) lines observed with ground-based telescopes,
we also derive a kinetic temperature that is definitely above 300 K for the outflowing
gas. 
In this case, caution should be taken since we are abl eto trace different
gas components, as suggested by CH$_3$OH, the gas at higher excitation 
being traced by CS(12--11).
If we analyse only the (2--1)/(3--2) intensity ratio, the non-LTE approach does
not allow us to constrain
the temperature in this way, but we are able to infer $n_{\rm H_2}$ of around 4 $\times$ 10$^4$ cm$^{-3}$.
Interestingly, when we check for a possible dependence of $n_{\rm H_2}$ on velocity,
the LV range is found to be indicative of a denser medium ($\sim$ 10$^5$ cm$^{-3}$) by
an order of magnitude with respect to the MV gas.

\section{Conclusions}

We have presented the HIFI unbiased spectral survey in the 555-636 GHz
band towards the bright bow-shock B1 of the L1157 protostellar outflow. 
For the first time, we have detected high-excitation (up to
$\simeq$ 200 K) emission lines of species whose abundance is
largely enhanced in shocked regions (e.g., H$_2$O, NH$_3$, H$_2$CO,
CH$_3$OH). This has allowed us to trace with these species  
the existence of a high excitation component with $T_{\rm kin}$
$\ge$ 200--300 K. 
Temperature components from $\sim$ 300 K to $\sim$ 1400 K have been inferred 
from the analysis of the H$_2$ pure rotational lines (Nisini et al., in prep.).
Therefore the present observations provide a link
between the
gas at $T_{\rm kin}$ 60--200 K previously observed from the ground and 
the warmer gas probed by the H$_2$ lines.
We plan to perform additional HIFI observations in the THz region 
towards L1157-B1 
to observe more species and transitions, thus to be able to derive reliable abundances and 
study of the different gas components associated with the bow structure.

\begin{acknowledgements}
HIFI has been designed and built by a consortium of institutes and
university departments from across
Europe, Canada and the United States under the leadership of SRON Netherlands Institute
for Space Research, Groningen, The Netherlands and with major contributions from Germany, France
and the US. Consortium members are: Canada: CSA, U.Waterloo; France: CESR, LAB, LERMA, IRAM;
Germany: KOSMA, MPIfR, MPS; Ireland, NUI Maynooth; Italy: ASI, IFSI-INAF, Osservatorio
Astrofisico di Arcetri-INAF; Netherlands: SRON, TUD; Poland: CAMK, CBK; Spain:
Observatorio Astron\'omico Nacional (IGN),
Centro de Astrobiolog\'{\i}a (CSIC-INTA). Sweden: Chalmers University of Technology -
MC2, RSS \& GARD;
Onsala Space Observatory; Swedish National Space Board, Stockholm University -
Stockholm Observatory;
Switzerland: ETH Zurich, FHNW; USA: Caltech, JPL, NHSC.
We thank many funding agencies for financial support.
\end{acknowledgements}


\begin{thebibliography}{}

\bibitem[1997]{bp97}
Bachiller R., \& Per\'ez Guti\'errez M. 1999, ApJ, 487, L93 (BP97)
\bibitem[1995]{bach95}
Bachiller R., Liechti S., Walmsley C.M., \& Colomer F. 1995, A\&A, 295, L51
\bibitem[2001]{bach01}
Bachiller R., Per\'ez Guti\'errez M., Kumar M.S.N., \& Tafalla M. 2001, A\&A 372, 899
\bibitem[2007]{milena07}
Benedettini M., Viti S., Codella C., et al. 2007, MNRAS, 381, 1127 (BVC07) 
\bibitem[2009]{code09}
Codella C., Benedettini M., Beltr\'an M.T., et al. 2009, A\&A, 507, L25
\bibitem[1995]{davis}
Davis C.J., \&  Eisl\"offel J. 1995, A\&A 300, 851
\bibitem[2010]{hifi}
de Graauw Th., et al. 2010, this volume 
\bibitem[1998]{garay}
Garay G., K\"ohnenkamp I., Bourke T.L., Rodr\'{\i}guez L.F., \& Lehtinen K.K. 1998, ApJ, 509, 768
\bibitem[1999]{gold}
Goldsmith P.F. \& Langer W.D. 1999, ApJ, 517, 209 
\bibitem[1996]{gueth96}
Gueth F., Guilloteau S., \& Bachiller R. 1996, A\&A 307, 891
\bibitem[1998]{fred98}
Gueth F., Guilloteau S., \& Bachiller R. 1998, A\&A, 333, 287
\bibitem[2008]{gus08b}
Gusdorf A., Pineau Des For\^ets G., Cabrit S., \& Flower D.R. 2008, A\&A 490, 695
\bibitem[2008]{jimenez}
Jim\'enez-Serra I., Caselli P., Mart\'{\i}n-Pintado J., \& Hartquist T.W. 2008, A\&A 482, 549
\bibitem[2007]{jorge}
J\o{}rgensen J.K., Bourke T.L., \& Myers P.C. 2007, ApJ, 659, 479
\bibitem[2010]{letter2}
Lefloch B., Cabrit S., Codella C., et al. 2010, this volume
\bibitem[2007]{looney}
Looney L.W., Tobin J.J., \& Kwon W. 2007, ApJ, 670, L131 
\bibitem[2005]{muller}
M\"uller H.S.P., Sch\"oier F.L., Stutzki J., \& Winnewisser G. 2005, J.Mol.Struct., 742, 215 
\bibitem[2009]{neufeld09}
Neufeld D.A., Nisini B., Giannini T., et al. 2009, ApJ, 706, N09
\bibitem[2007]{nisini07}
Nisini B., Codella C., Giannini T., et al. 2007, A\&A, 462, 163 
\bibitem[1998]{pickett}
Pickett H.M., Poynter R.L., Cohen E.A.,  Delitsky M.L., Pearson J.C., and M\"uller H.S.P. 1998,
J. Quant. Spectrosc. \& Rad. Transfer, 60, 883
\bibitem[2010]{herschel}
Pilbratt G., et al. 2010, this volume
\bibitem[2004]{pottage}
Pottage J.T., Flower D.R., \& Davis S.L. 2004, MNRAS, 352, 39
\bibitem[2005]{lamda}
Sch\"oier F.L., van der Tak FFS., van Dishoeck E.F., \& Black J.H. 2005, A\&A, 432, 369
\bibitem[1995]{tafabachi}
Tafalla M., \& Bachiller R. 1995, ApJ 443, L37
\bibitem[1969]{viotti}
Viotti N.R. 1969, Mem. Soc. Astron. Ital., 40, 75
\bibitem[1998]{vanblake}
van Dishoeck E.F., \& Blake G.A. 1998, ARA\&A, 36, 317
\bibitem[2007]{radex}
Van der Tak F.F.S., Black J.H., Sch\"oier F.L., Jansen D.J., \& van Dishoeck, E.F. 2007, A\&A, 468, 627

\end{thebibliography}
\end{document}